**Photon-assisted ultrafast electron-hole plasma expansion in direct band semiconductors**

T. Troha,[1] F. Klimovič,[2] T. Ostatnický,[2] H. Němec,[1] P. Kužel,[1,*]

[1] FZU - Institute of Physics of the Czech Academy of Sciences, Na Slovance 2, 182 00 Prague 8, Czech Republic
[2]Charles University, Faculty of Mathematics and Physics, Ke Karlovu 3, 121 16 Prague 2, Czech Republic

*e-mail of corresponding author: kuzelp@fzu.cz

**Abstract**: Time-resolved terahertz spectroscopy is used to investigate formation and ultrafast long-distance propagation of electron-hole plasma in strongly photoexcited GaAs and InP. The observed phenomena involve fundamental interactions of electron-hole system with light, which manifest themselves in two different regimes: a coherent one with the plasma propagation speeds up to $c/10$ (in GaAs at 20 K) and an incoherent one reaching up to $c/25$ (in InP at 20 K), both over a macroscopic distance $> 100\ \mu\mathrm{m}$. We explore a broad range of experimental conditions by investigating the two materials, by tuning their band gap with temperature and by controlling the interaction strength with the optical pump fluence. Our interpretation suggests that the observed phenomena should occur in most direct band semiconductors upon strong photoexcitation with low excess energy.

## 1. Introduction

Semiconductors are still the most prominent materials for the conception and development of modern electronic and opto-electronic components; therefore, they belong to the best understood solid materials ever. Nevertheless, surprising new phenomena are still being discovered even in off-the-shelf semiconductor wafers and by using relatively simple table-top spectroscopic experiments.

The transport of (photoexcited) conduction-band carriers in semiconductors is a fundamental process in electronic and opto-electronic applications. Ballistic transport regime is considered the fastest mechanism of the charge carrier transport [1]; its speed (carrier group velocity $v_g$) is, however, limited by the band structure ($v_g = \partial E/\partial k$) and it does not exceed $(1-2)\times 10^6\ \mathrm{m/s}$ in any known crystalline semiconductor. At high photoexcited carrier densities various fast electronic transport mechanisms have been observed, including Fermi pressure driven electron-hole plasma (EHP) expansion [2,3,4], thermodiffusion [5], screening of electron-phonon interaction [6] and processes involving stimulated absorption and emission of radiation [7,8]. However, those works only reported the EHP expansion rates below the limit imposed by the band structure. Recently, we have shown that GaAs submitted to strong femtosecond excitation with wavelength slightly exceeding the band gap leads to unexpected phenomena [9]. Namely, we observed stimulated emission of photons with the energy close to the bandgap energy, which manifests itself as an ultrafast transfer of electron-hole pairs (with speeds of a few tenths of *c*) over extremely large distances (tens to hundreds of micrometers).

Terahertz (THz) radiation strongly interacts with free charges and can be used to determine the extent and time evolution of EHP in a semiconductor [10,11]. Indeed, surfaces of a layer of dense EHP behave as metallic-like (i.e, highly reflecting) mirrors in the THz range, and the position of the interface between EHP and unexcited material can be measured in a setup for time-resolved THz spectroscopy. This method employs coherent picosecond pulses of THz radiation which are detected using a phase-sensitive technique. This allows one to carry out a time-of-flight variant [10,11,12] of optical pump – THz probe measurements with femtosecond resolution. The position of EHP front can be thus determined with sub-micrometer accuracy.

In this work, we concentrate on the question whether other materials, besides bulk GaAs [9], can reveal ultrafast propagation of the EHP front and what are the critical parameters for such observation. We ask this question since the ultrafast EHP expansion has not been reported previously despite the fact that we use the common pump and probe geometry. We demonstrate that the ultrafast EHP kinetics can be triggered also in a bulk InP. We argue that this EHP dynamics is quite a general phenomenon, which is likely to occur in most direct-bandgap semiconductors, provided the femtosecond laser pulse excitation complies with the following conditions: (*i*) the photoexcitation wavelength ensures an excess energy of photoexcited charge carriers of the order of a few hundred meV or smaller and (*ii*) a sufficient photon fluence is used to induce absorption bleaching in a layer at least a few micrometers thick. Upon tuning the excess energy (and also in relation with the carrier cooling rate) incoherent or coherent regime of the plasma propagation is observed characterized by a stimulated emission of radiation or coherent Rabi dynamics, respectively. We describe the crossover between these two regimes.

## 2. Materials and methods

*Experimental setup*

Our experimental setup (Fig. 1) is based on a Ti-sapphire amplified laser system (Spitfire ACE) with 800 nm central wavelength, 40 fs pulse length, 1 mJ pulse energy, and 5 kHz repetition rate. The pulse train is divided into the pump, THz and sampling branch. The pump pulse (propagating colinearly with the THz probe pulse) is used for the generation of the electron-hole plasma (EHP) at the front surface of the investigated wafer. The THz probe pulse is generated by optical rectification and detected via electro-optic sampling in a pair of (110) 1 mm thick ZnTe crystals.

We use a time-of-flight variant of the optical pump – THz probe spectroscopy [10]. The inner surface of the photoexcited EHP acts as a metallic-like mirror; the thickness of the EHP layer is determined from the arrival time of the pulse to the electro-optic sensor. In short, the THz pulse enters the semiconductor slab at its front side before optical excitation. Subsequently, the pump pulse impinges on the front side of the sample, where it is absorbed and generates an EHP with initial thickness $l_0$. The THz pulse is partially reflected at the rear side: the directly transmitted pulse serves as a reference of experimental timing while the internally reflected pulse probes the plasma-front state inside the sample. The time of the arrival of the probe pulse to the electro-optic detector carries information about the actual position of the EHP edge. By changing the delay between the pump and the probe pulse we are able to determine the course of EHP expansion. If the probe pulse waveform is not substantially reshaped, then the accuracy of the time shift of the detected probe pulse is estimated to $\sim 5$ fs (this corresponds to the plasma thickness uncertainty of ~0.4 μm).

The semiconductor slab was attached behind a metallic aperture with a diameter of 2 mm. The photon fluences of excitation pulses used in the experiments were in the range $(0.08 - 2.4) \times 10^{16}\ \mathrm{cm}^{-2}$, high enough to saturate absorption close to the front face of the wafer and thus generate a layer of EHP with fluence-dependent initial thickness.

The spatial and temporal properties of the pump pulse can affect several processes of the studied phenomenon. In particular, its duration controls the two-photon absorption (TPA), and its spatial profile determines the distribution of free carriers and thus the efficiency of the stimulated emission. Whereas uncertainties in these characteristics do not affect any qualitative conclusion, they can influence to some extent, e.g., the observed EHP expansion rate. The measurements presented throughout this paper were performed with particular settings of the laser and of the beam path. However, due to the above reasons, the measured values might slightly differ from those observed previously [9,10].

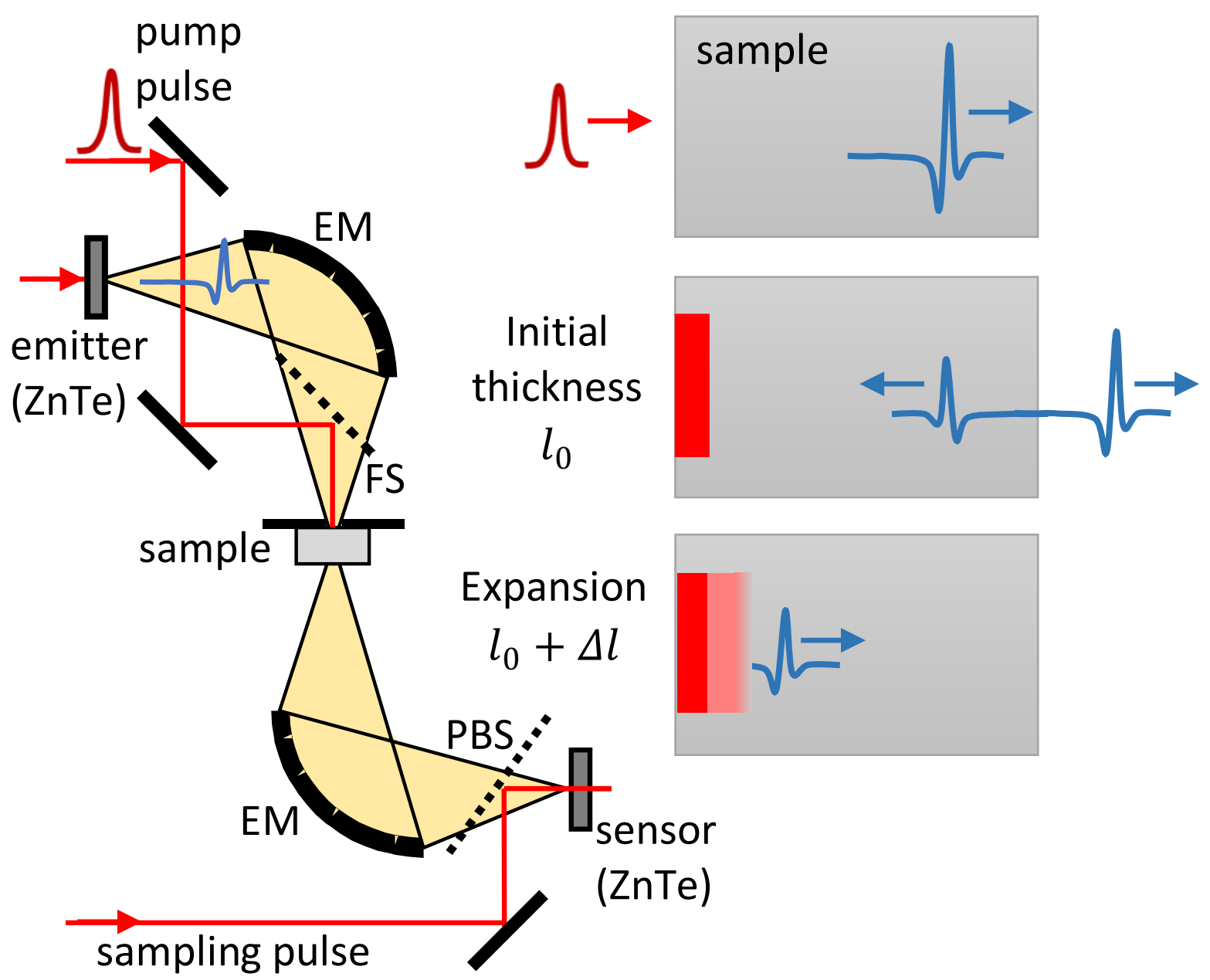

Figure 1: Scheme of the time-of-flight variant of the optical pump – THz probe experiment. EM: ellipsoidal mirrors, PBS: pellicle beam-splitter, FS: fused silica beam-splitter with high-reflective dielectric coating for 800 nm. The detail of the pulse propagation in the sample is shown at the right-hand side. The pump pulse excites the front surface of the sample and THz pulse probes the EHP expansion by its internal reflection at the photoexcited surface.

*Samples*

We used samples from common wafers of high-resistivity GaAs and high-resistivity InP with optically polished surfaces (exhibiting standard electron mobilities). The band structures of these direct-gap materials are qualitatively similar (relevant parameters are summarized in Table 1). The excess energy of an electron-hole pair can be tuned from 30 to 130 meV in GaAs, and from 130 to 200 meV in InP as a result of the band gap shrinkage between 20 K and 300 K. This approach thus enables us to directly compare two materials with identical carrier excess energy (although at different lattice temperatures). The energy difference between the split-off band and the valence band in InP is significantly smaller than in GaAs and, consequently, the split-off band can contribute to the absorption of the pump pulse in InP as pointed out in Table 1. The number of states per unit volume estimated in the table corresponds to the number of states accessible by direct optical excitation using 40 fs pulses centered at 800 nm; the bleaching occurs when one half of these states is filled. The number of accessible states in GaAs at 300 K and in InP at 20 K is similar (but not exactly equal, due to the slight difference of effective electron masses and due to the contribution of the split-off band in InP). While the TPA coefficient at 800 nm has been determined for GaAs at room temperature [10], it has been reported neither for GaAs at low temperatures, nor for InP at any temperature. The TPA coefficient depends on the joint density of states involved in the two-photon absorption, which should correlate with the joint density of states of the single photon absorption (SPA) at a double frequency. The latter is directly proportional to the imaginary part of the linear permittivity which is known for GaAs and InP [13]. The correlation between the TPA and SPA is likely to be similar for the two binary semiconductors with similar band-structures. Based on these simple arguments, we expect that the TPA in InP could be somewhat weaker than in GaAs but of the same order of magnitude.

| **Properties at 300 K** | **GaAs** | **InP** |
|---|---|---|
| Bandgap (eV) | 1.42 | 1.35 |
| Excess energy* (eV) | 0.13 | 0.20 |
| Split-off band (eV) | -0.34 | -0.11 |
| *L*-valley (eV) | 1.81 | 2.01 |
| Electron mass | 0.067 | 0.08 |
| Heavy-hole mass | 0.5 | 0.6 |
| Light-hole mass | 0.08 | 0.09 |
| Split-off hole mass | 0.17 | 0.21 |
| Number of states ($10^{18}$ $\mathrm{cm}^{-3}$)** | 5 | 8 (without SOB)<br>13 (with SOB) |
| THz refractive index | 3.60 | 3.55 |
| **Properties at 20 K** | **GaAs** | **InP** |
| Bandgap (eV) | 1.52 | 1.42 |
| Excess energy* (eV) | 0.03 | 0.13 |
| Number of states ($10^{18}$ $\mathrm{cm}^{-3}$)** | 2 | 6 (without SOB)<br>8 (with SOB) |

Table 1: Important properties of GaAs and InP at room temperature (upper part of the table) and at 20 K (lower part of the table). From the review paper [14]. *Excess energy of electrons generated by 800 nm pulses. **Estimated as number of electron states per unit volume that can be reached by optical excitation with 40 fs pulses at 800 nm, excluding or including the split-off band (SOB).

## 3. Results

The positions of the excited/unexcited semiconductor interface were measured as a function of the pump-probe delay. The results were deduced from the time-domain advancements $\Delta t$ of the measured probing waveforms, namely we read the time of the zero-crossing of the electric field between their first two pronounced extrema (for details see Supplementary material of [9]). The spatial position $\Delta l$ of the EHP front is then calculated as $\Delta l = c\Delta t/2n_{\mathrm{THz}}$, where $n_{\mathrm{THz}}$ stands for the steady-state refractive index of the sample and $c$ is the speed of light in vacuum. As a reference (zero thickness) we take the reflection on the air-semiconductor interface obtained when the pump pulse impinges on the sample significantly after the THz probe pulse. In this way we eliminate a possible role of the heating of the sample on the optical path length inside the semiconductor with the pump beam on and off. Note that we also particularly avoid the response generated by pre-pulses in the pump beam: a pre-pulse, even if relatively weak, may be able to generate a non-negligible carrier density for the high pump fluences used in this work. The surface photoexcited by a weak pulse does not act as a good metallic mirror (its reflectivity is considerably lower than 1) and therefore can lead to a complex reshaping of the reflected THz pulse.

The zero pump-probe delay was chosen close to the point at which the THz pulse probes the surface of the wafer just starting to be influenced by the pump. Although this time can be chosen arbitrarily to some extent due to the picosecond length of the THz pulse, the relative timing among the measured waveforms is unambiguous.

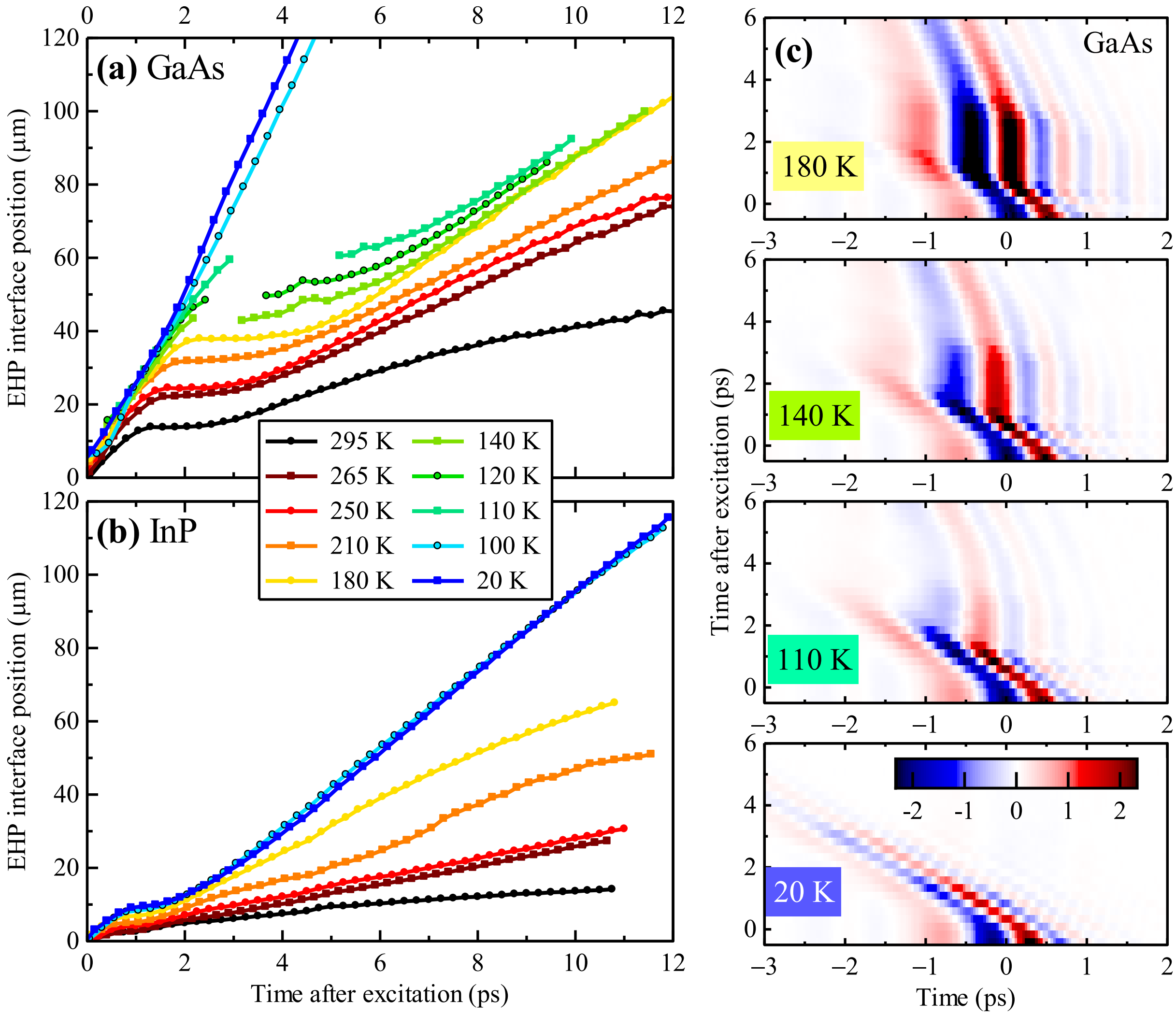

**Figure 2**: Temperature dependence of the EHP extent in GaAs (a) and InP (b) for the pump fluence of $1.7 \times 10^{16}$ photons/cm$^2$. The legend showing the temperatures is the same for both samples. (c) Colormap images of the measured THz waveforms reflected from the EHP as a function of the pump–probe delay (time after excitation) for GaAs at selected temperatures. The color scale in (c) is in arbitrary units, but the same for all the panels.

In Fig. 2(a,b) we present the temperature dependence of the position of the plasma front for GaAs and InP illuminated at a high pump photon fluence of $1.7 \times 10^{16}$ cm$^{-2}$. We can distinguish two regimes of the EHP behavior. The first regime is observed in GaAs at high temperatures ($> 140$ K) and at all the measured temperatures in InP. In this case, an initial increase of the plasma extent (during $\sim 0.5 - 1.5$ ps) is always followed by a plateau (also $\sim 0.5 - 1.5$ ps) corresponding to the carrier relaxation (cooling) to the bottom of the conduction band. After these phases, the plasma expansion occurs through stimulated emission and reabsorption of light with the wavelength close to the bandgap edge. We observe that plasma expands faster in both materials upon the temperature decrease.

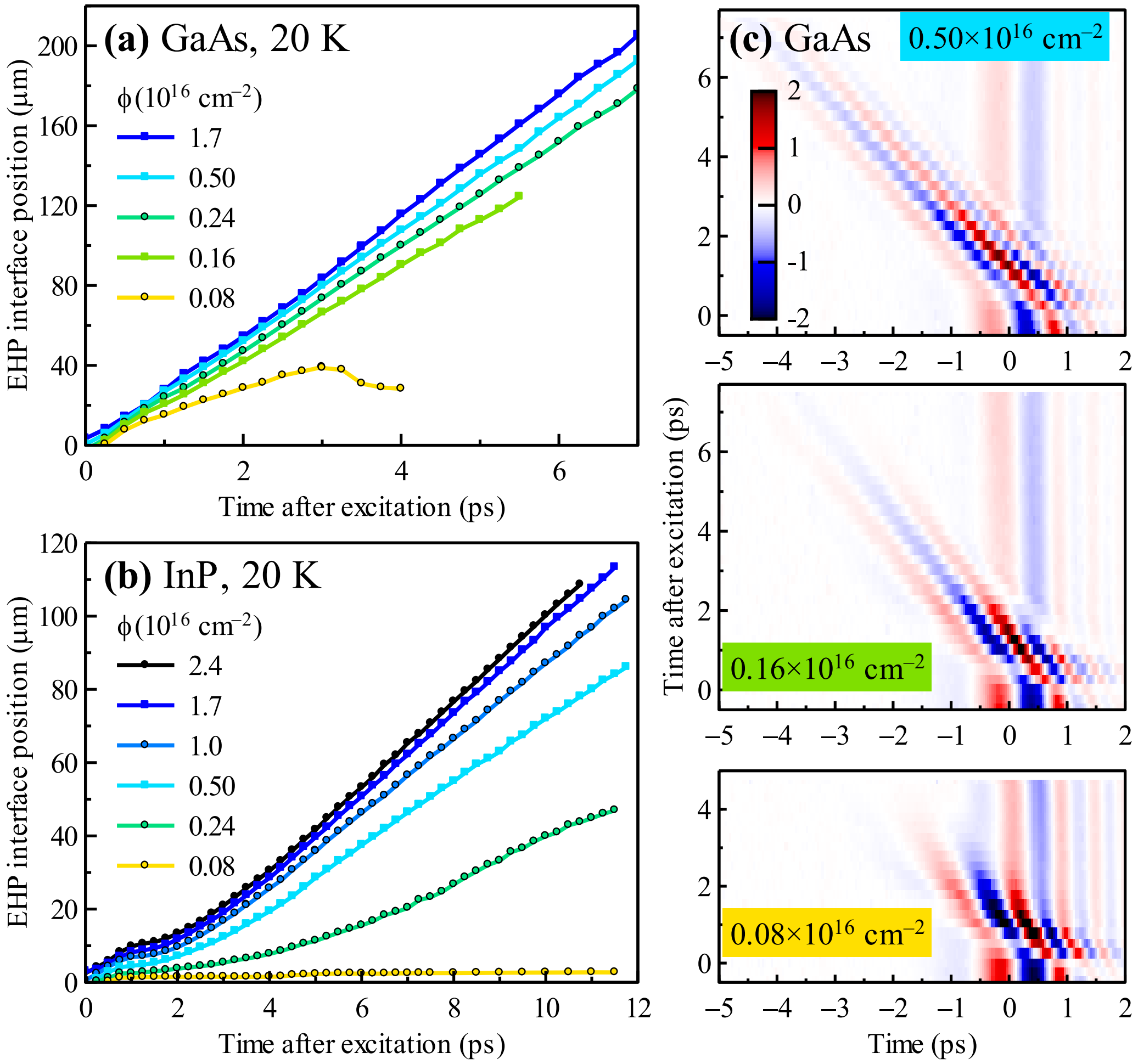

**Figure 3**: Fluence dependence of the photoexcited interface propagation in GaAs (a) and InP (b) at 20 K. (c) Colormap images of the measured THz waveforms reflected from the EHP as a function of the pump–probe delay (time after excitation) for GaAs at selected fluences. The color scale in (c) is the same for all the panels. The oblique signal represents a part of the THz wave form reflected at the soliton pulse while the vertical lines at large pump–probe delays correspond to the signal reflected on the front surface of GaAs.

The second regime occurs in GaAs only at low temperatures (≤ 100 K). Namely, a soliton propagation is observed due to coherent electron-photon interaction and Rabi dynamics as described in [9]. These coherent phenomena do not involve the carrier energy relaxation (cooling) plateau and the reflection of the THz pulse occurs, in fact, on a plasma sheet with the thickness defined by the soliton width in the propagation direction. This regime is not detected in InP. A crossover between these two regimes is observed in GaAs at intermediate temperatures (110–140 K). It is accompanied by reshaping of the THz waveforms: in this particular situation, the zero-crossing of the THz field reflects the changes in the waveform shape rather than the EHP interface position. For this reason, such data points are omitted in Fig. 2(a) and in turn, the curves for temperatures 110 – 140 K are not continuous. To illustrate better this crossover, we thus show a set of measured waveforms at several temperatures as colormap plots in Fig. 2(c). For 180 K (top panel) we observe a clear plateau in the waveform shift near the pump–probe delay of 2 ps. At intermediate temperatures (140 and 110 K, middle panels) a competition is observed between the energy relaxation of carriers contributing to the plateau phase and the formation of a (short-lived) pulse exhibiting the coherent Rabi dynamics; clearly, this pulse propagates over a longer distance at 110 K than at 140 K. Soliton propagation only is observed at 20 K (bottom panel).

The fluence dependence of the photoexcited interface propagation at 20 K is shown in Fig. 3 both for GaAs and InP. The data for GaAs, plotted in Figs. 3(a) and (c), show quite a complex behavior. The soliton propagation is observed down to the excitation photon fluence of $0.16 \times 10^{16}\ \mathrm{cm}^{-2}$ and its speed slightly decreases with the decreasing fluence. However, since the photon energy is very close to the band edge, the photo-generated EHP inside the weakening pulse does not represent a metallic-like mirror for the THz radiation anymore; therefore a part of the THz pulse is reflected on the front surface of the sample, see Fig. 3(c). As observed in the middle panel of Fig. 3(c), for the fluence of $0.16 \times 10^{16}\ \mathrm{cm}^{-2}$, the two reflections have a comparable intensity. For even lower pump fluences the soliton propagation stops completely as observed in the bottom panel of Fig. 3(c). On the other hand, in InP at 20 K we observe that the plasma expansion starts always after some waiting time (plateau), see Fig. 3(b). Upon a decrease of the pump fluence the plasma expands slower in InP until it stops completely at the fluence of $0.08 \times 10^{16}\ \mathrm{cm}^{-2}$.

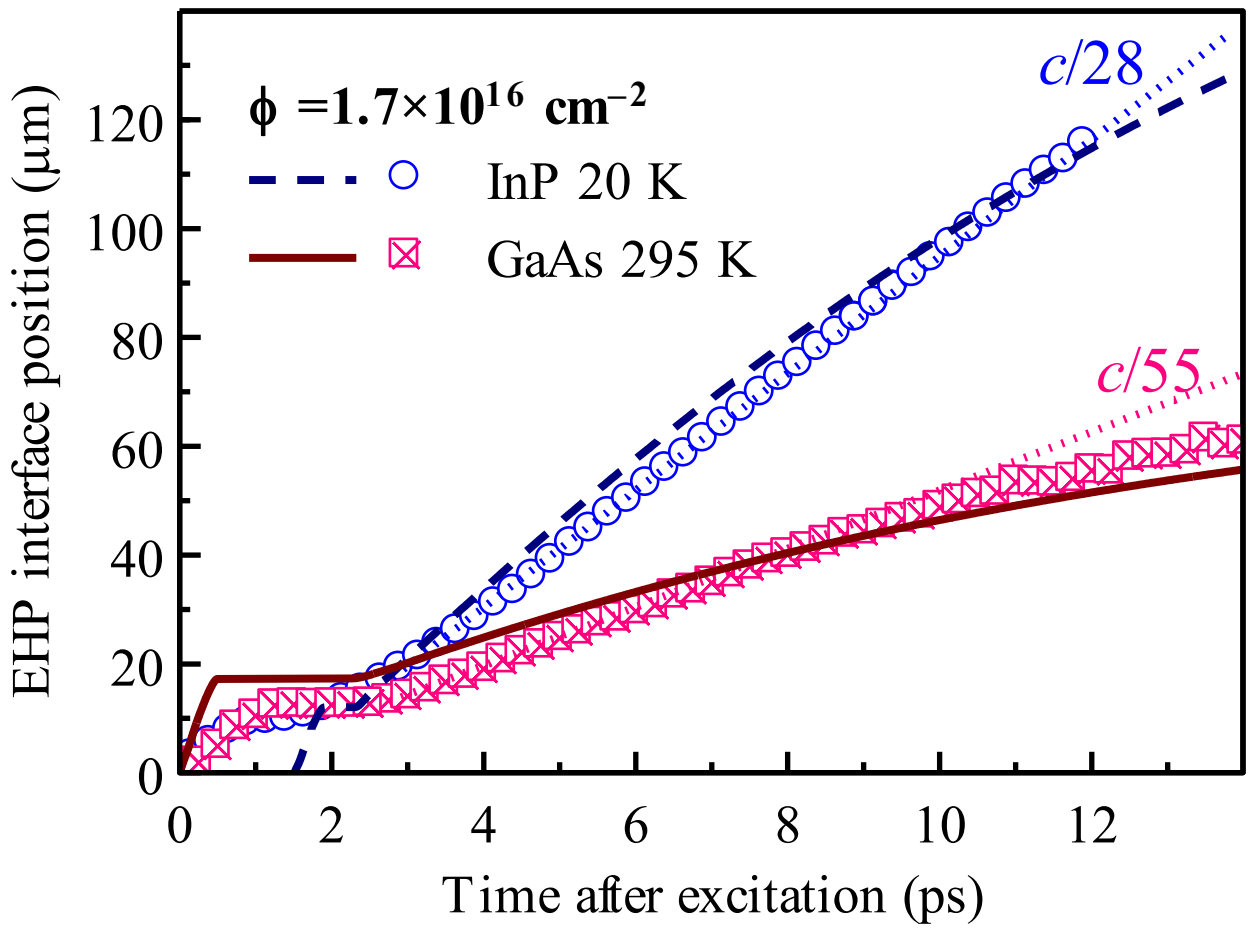

**Figure 4**: Comparison of the EHP interface position as a function of the pump probe delay for GaAs at room temperature and InP at 20 K. The two materials have the same band gap energy at these temperatures. Pump photon fluence: $1.7 \times 10^{16}\ \mathrm{cm}^{-2}$ in both cases. Symbols: experimental data, dotted lines: slopes of experimental data (corresponding EHP expansion speed is indicated), solid and dashed line: results of the calculations for GaAs and InP, respectively, within the model for incoherent regime described in Sec. 4. For InP, the model dynamics predicts a shorter plateau phase than observed; to be able to directly compare the expansion phase, we shifted the theoretical curve for InP in time by 1.5 ps to the right in order to match the expansion onset with that observed in the experiment.

In Fig. 4 we directly compare the EHP expansion dynamics in GaAs and InP for the same excess energy, i.e., in GaAs at room temperature and in InP at 20 K. The plasma front position in both cases shows a plateau (shorter for InP) and an expansion which is about twice faster in InP than in GaAs. This is a rather surprising result taking into account the similarity of the band structure of the two materials.

## 4. Discussion

*Overview of the theoretical description*

In [9] two regimes were proposed to explain the temperature dependence of the observed phenomena in GaAs: an incoherent regime valid at high temperatures and a coherent regime governing the process at low temperatures. We considered two photon fields (incident pulse with frequency $\omega_1$ and a re-emitted radiation with frequency $\omega_2$) and two relevant energy levels which participate in the system dynamics. The upper level 1 ($N_1$ is the number of available states per unit volume) is populated by the direct single-photon absorption of the pump laser field ($\omega_1$).

In the incoherent regime the level 1 plays a role of the reservoir for the bottom level 2 with a lower number of states per unit volume ($N_2$), which is the source of an amplified stimulated emission of "recycled" photons ($\omega_2$). These photons leave the originally excited EHP region, become absorbed in the unexcited part of the crystal and contribute to the plasma edge shift in space. The initial $\sim$ ps long plateau, which is always experimentally observed at the beginning of the EHP expansion dynamics, then reflects the build-up of the population inversion at the level $N_2$, i.e., a relaxation of carriers from $N_1$ to $N_2$. The coherence of the system is instantaneously lost during this relaxation process.

In the coherent regime the two photon fields overlap in frequency, no significant energy relaxation occurs, and the coherent electron-photon interaction leads to the Rabi dynamics connected to a formation of solitons carrying the EHP plasma deep into the sample.

*Coherent regime*

The coherent regime takes place when three conditions are met: the photon fields $\omega_1$ and $\omega_2$ overlap (within the excitation pulse bandwidth) in frequency, the carrier decoherence time exceeds the pump pulse length and the pump pulse fluence is sufficient to produce the so called pulse area (i.e., a change in the phase of the population inversion oscillation) of at least $2\pi$, which causes just one Rabi flop in the population inversion [15,16,17,18,19]. Such coherent electron-photon interaction develops in GaAs at low temperatures; under these conditions, the reflection of the THz pulse monitors the long-distance propagation of a $2\pi$ soliton pulse with the speed [20,21]:

$$u = \frac{c}{n_g + \frac{1}{2}c\alpha t_p}, \tag{1}$$

where $n_g$ is the group refractive index and $\alpha$ is the (linear) absorption coefficient in the semiconductor, and $t_p$ is the pump pulse duration.

The coherent regime is demonstrated in GaAs at low temperatures. Indeed, at 20 K and with fluence $\phi = 1.7 \times 10^{16}\ \mathrm{cm}^{-2}$ the pulse area is, following our estimations, $4 \times 2\pi$ [9], which means that several copropagating solitons can be formed each with a spatial thickness of a few micrometers. The THz reflectance is then given by the sheet conductivity of the whole soliton structure. At this fluence, the sheet conductivity is high enough to efficiently reflect the THz radiation as observed in the bottom panel of Fig. 2(c). Upon a decrease of the pump fluence the pulse area decreases with the square root of the excitation fluence. The cut-off fluence below which even a single soliton pulse is not really formed is then $\sim 0.1 \times 10^{16}\ \mathrm{cm}^{-2}$. This is observed in Fig. 3(c), where the reflection on the soliton structure becomes progressively less intense as the pump fluence is diminished and, finally, in the bottom panel, the soliton does not propagate. At the same time, we observe a reflection on the photoexcited front surface of the wafer; however, we do not observe any propagation of this interface since the density of carriers is not sufficiently high. Similar situation is observed if we heat the sample, Fig. 2(c); indeed, as the temperature is increased, the bandgap shrinks and the spectral overlap between the photon fields $\omega_1$ and $\omega_2$ becomes progressively smaller. Less carriers can then participate to the coherent interaction and the pulse related to the Rabi dynamics is weak (110 K) or cannot be formed anymore (180 K). At these temperatures a reflection at the EHP close to the GaAs surface is also observed and, since the overall photon fluence is high, the incoherent regime takes over and the EHP expansion restarts at later times.

Relatively slow propagation of the excitation pulse (i.e., at a speed smaller than expected $c/n_g$) is observed also in the initial phase of the EHP build-up before the appearance of the plateau in the dynamics for both GaAs and InP. To address this effect, we estimate the time required by light to propagate through (and thus form) the initial plasma thickness $l_0$: $l_0 n_g/c + t_1 + t_2$. The first term $\lesssim 150$ fs in GaAs at room temperature, $t_1$ is a characteristic time-domain width of the optical pump pulse, during which the light intensity is high enough to saturate the absorption ($\lesssim 80$ fs in our case) and $t_2$ is related to the so-called depletion time (time during which the generated plasma fills out an exponentially decaying spatial distribution) [22]. This time can be significant in materials with a long linear penetration depth (e.g., Si excited at 800 nm [11,22]). However, in our case of GaAs or InP with a sub-micrometer linear penetration depth it remains quite short ($\lesssim 45$ fs), since the initial plasma propagation is essentially due to the saturation of single photon absorption. The predicted initial plasma formation time is then $\lesssim 300$ fs, whereas the corresponding observed initial increase of the plasma extent lasts more than a picosecond. The reason for this behavior can be twofold. (*i*) A reshaping of the picosecond THz pulse is observed during this initial stage since different parts of the THz pulse find the close neighborhood of the EHP front in a different (and non-stationary) state. In other words, we observe a "convolution" of the fast plasma propagation with the two adjacent plateaus (one at negative times before the carriers are generated and the other due to the carrier energy relaxation) leading to a lowering of the apparent speed. (*ii*) A possible short-lived coherent regime might occur within the directly excited electronic states well above the bandgap (photon field $\omega_1$) leading to a slower propagation of the optical pump pulse during the absorption process. Such regime would decay rapidly due to the fast momentum relaxation of the charges. An existence of this last contribution is further supported by the dynamics observed in GaAs shown in Fig. 2(a). The initial absorption phase becomes progressively longer on cooling from the room temperature while the observed speed (slope of the curves) remains approximately constant and comparable to the soliton speed detected at low temperatures.

For InP in the entire temperature range, the excess energy exceeds the bandwidth of the excitation pulse as well as the optical phonon energy. Under these conditions, a rapid carrier energy relaxation occurs which prevents the development of the coherent regime for a long time. Indeed, for InP we always observe the plateau which is characteristic for the incoherent regime [Figs. 2(b) and 3(b)].

*Incoherent regime*

Although the band structures of GaAs and InP are very similar, well pronounced differences of the plasma expansion in incoherent regime have been observed between the two materials. First, in InP the incoherent process is observed down to 20 K and its EHP propagation speed finally saturates upon decreasing the temperature. In contrast, the EHP dynamics is governed by the soliton propagation in GaAs at 20 K. Second, EHP expansion speeds differ by a factor of 2 between GaAs (295 K) and InP (20 K) with the same excitation excess energy (Fig. 4). These findings led us to an extension of the kinetic model compared to our original model in [9].

Each electronic state can absorb/amplify electromagnetic radiation, depending on the actual population of this state (more precisely, depending on the inversion of the pair of electromagnetically coupled electron and hole). In a real experiment, the relaxation kinetics and state depletion by photon emission leads to a space-dependent dynamics in which many states within the band dispersion emit and reabsorb photons in a complex cascade. This kinetics, despite its complexity, can be effectively described in terms of an "active level" with the number of states per unit volume $N_2$, fed by a "reservoir" with the number of states per unit volume $N_1$. Our kinetic equations are shown in the Appendix. The essential extension of the model is an explicit inclusion of the $L$-valley carriers ($n_L$), which are generated through TPA and subsequently relax to the reservoir. Furthermore, we deeply rethought the interpretation of the degeneracy of level 2, $N_2$, which reveals to be the crucial parameter for the observed dynamics and its temperature dependence. Indeed, following the estimate of the early time plasma expansion speed $u$, Eq. (5) in [9]:

$$u = \frac{l_0}{2\tau}\frac{(N_1 - N_2)}{N_2}, \tag{2}$$

we find that $u$ may acquire large values when $N_2$ is relatively small ($l_0$ is the initial thickness of EHP, $\tau$ is the mean relaxation time from level 1 to level 2 and represents the inverse intra-band cooling rate of electrons). This counter-intuitive behavior can be better understood when a real system with a continuum of levels is considered. The distinct energy levels are characterized by different absorption coefficients (due to their different occupancy and degeneracy) and, therefore, by a distribution of expansion speeds. In extremis, photons emitted by electrons with an infinitely small excess energy above the band gap travel through the semiconductor practically unabsorbed, thus causing an infinitely small EHP density to expand at the speed of light in the material. In other words, the experimentally observed motion of the plasma front edge is determined by the fastest speed $u(N_2)$ given by (2), where the lowest acceptable value of $N_2$ satisfies some physical constraints discussed below.

Our further discussion of the meaning and interpretation of $N_2$ is based on the scheme shown in Fig. 5. First we discuss the constraints in the plasma expansion region (region B in Fig. 5) which impose a minimum (or critical) free electron density $n_{2,c}$ in the region B (the corresponding level degeneracy is then $N_{2,c} = 2n_{2,c}$). First, a sufficiently large concentration of electrons $n_{2,0}$ must be created in the region B to reflect the THz pulse, defining the first lower limit for the level 2 degeneracy: $N_{2,0} = 2n_{2,0}$. Second, the propagation of the EHP front is possible only if the optical absorption is saturated. Such situation occurs in the region B when one half of electronic states available for absorption of photons at a given frequency is filled (including those reached by thermal fluctuations). The minimum density of thermalized electrons $n_{2,T}$ which leads to the saturation of the optical absorption then defines the second lower limit for the degeneracy: $N_{2,T} = 2n_{2,T}$. The critical degeneracy $N_{2,c}$ (electron density $n_{2,c}$) is then the larger of the two values $N_{2,0}$ and $N_{2,T}$ ($n_{2,0}$ and $n_{2,T}$). $N_{2,0}$ is important at low lattice temperatures when thermal fluctuations are negligible; $N_{2,T}$ becomes dominant at higher temperatures. This implies that the number of available states at level 2, $N_{2,c}$, is in fact temperature dependent (and that it is also tightly bound to the sensitivity of the experimental technique).

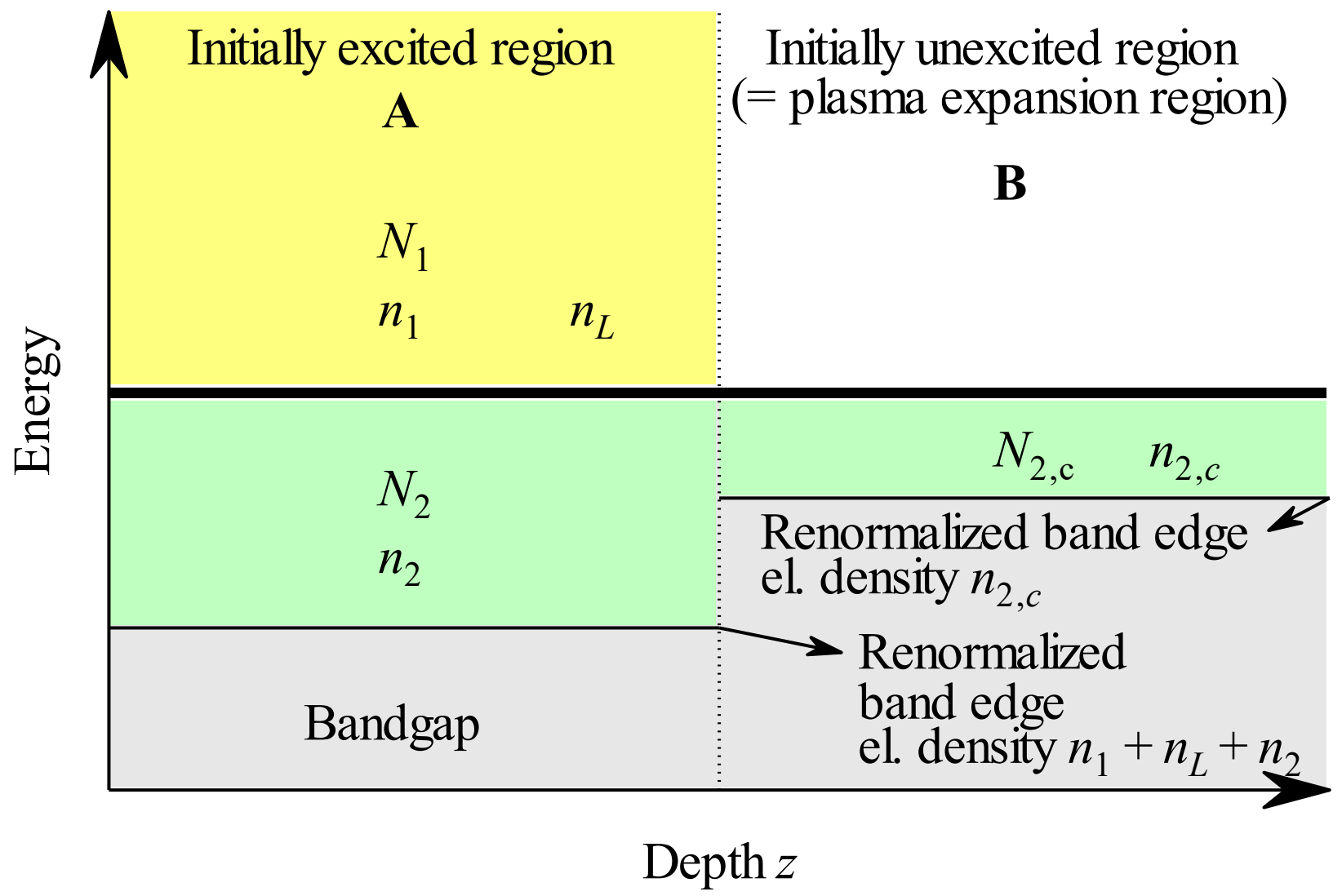

Figure 5: Scheme of the conduction band in space along the normal to the sample surface after its photoexcitation and during the plasma expansion. The initially excited region A is the part of the sample where the photon field at $\omega_1$ is absorbed; the plasma expands into the initially unexcited region B (see also [9]). The bottom of the conduction band is renormalized due to the high carrier concentration ($n_1 + n_L + n_2$ in region A, $n_{2,c}$ in region B). Level 1: yellow, level 2: green, bandgap: gray.

Now we turn our attention to the region A in Fig. 5. In fact, given the rather high carrier densities and small excess energies, the bandgap renormalization plays a non-negligible role in our estimations, since it is different in regions A and B due to the different carrier densities in there (Fig. 5). Consequently, the division energy line between levels 1 and 2 in the region A should be adjusted appropriately to ensure the critical electron density $n_{2,c}$ in the region B. This procedure allows us to define the final value $N_2$ to be injected into kinetic equations for the EHP expansion and allows us to describe the observed temperature variations of the expansion speeds. Technical details and deeper discussion of the evaluation of $N_2$ are provided in the Appendix.

Given these arguments, the difference between the dynamics of InP at 20 K and GaAs at 295 K becomes more apparent. Indeed, the evaluated degeneracies of level 2 differ among the two materials ($N_2 = 1.0 \times 10^{18}\ \mathrm{cm}^{-3}$ in InP at 20 K and $N_2 = 2.4 \times 10^{18}\ \mathrm{cm}^{-3}$ in GaAs at 295 K) and therefore the model predicts different speeds of expansion. In the comparison of the theoretical prediction and experimental results (Fig. 4) we observe a very good semiquantitative agreement of the data. We point out that the model predicts a significantly shorter phase of carrier energy relaxation (plateau phase) in InP than experimentally observed (for GaAs the length of the plateau phase is correct). To compare the EHP expansion dynamics we shifted the origin of the theoretical curve for InP in order to match the onset of the expansion. In this representation, it becomes apparent that the expansion dynamics is well reproduced by the model, which also predicts the correct difference of the expansion velocities in the two materials.

Fig. 6 shows a comparison of the temperature dependent plasma expansion in GaAs and InP and the corresponding results of the theoretical model for all the temperatures where the incoherent regime was observed. The overall agreement is very good. In particular, we find a semiquantitative accord of the temperature dependence of EHP expansion in GaAs. The agreement for InP is very good at low temperatures (up to about 180 K); at higher temperatures, the speed of expansion is overestimated by the theory to some extent (even if the decreasing trend of the expansion speed with increasing temperature is preserved). The energy relaxation (plateau) phase in InP is systematically underestimated; for this reason, theoretical curves were shifted in time in Fig. 6 to match the onset of expansion with the experiment.

The relaxation time $\tau$ was set to $2\ \mathrm{ps}$ in the numerical simulations; the value was selected in order to reproduce expansion speeds and trends in the experimental curves. We therefore conclude that the time $\tau$ is an effective cooling time of the whole electron-photon system, including the energy redistribution within the inhomogeneously excited plasma, provided by the cascade of photon emission and absorption.

Despite the complexity of this highly nonlinear problem and despite the number of degrees of freedom which can influence the quantitative results, the essence of the observed phenomenon is grasped quite well by the proposed simple model based just on effective populations of the "reservoir" and of the "active level" and relying on (mostly known) properties of the semiconductor band structures. The model then serves as a good quantitative estimate of the system behavior and illustrates well the underlying physical processes. The complex nature of the carrier relaxation is further indicated by the mentioned systematically shorter plateaus in InP; this suggests that the quasi-equilibrium state is established in a more complex process than the simple exponential decay represented by (A1–A5).

*Generalization*

The described ultrafast EHP expansion is closely related to the formation of a degenerate EHP and to an efficient stimulated emission and subsequent reabsorption. These particular phenomena should be in principle observable in the majority of common direct-gap semiconductors. The question of observing the ultrafast EHP expansion in other materials is thus much more about the laser pulse fluence and wavelength as discussed below.

The creation of a sufficiently thick degenerate EHP requires high fluence of the excitation field. The higher is the photon excess energy above the band gap, the larger is the excitation pulse attenuation by absorption due to the larger number of available state per unit volume and, in turn, the higher is the required excitation fluence. The highest possible fluence is, however, limited by the available laser technology and also by the damage threshold of the given material (e.g., the highest fluences used in our experiments with GaAs at 800 nm were still safely below its damage threshold [23,24]).

There may be slight quantitative (not qualitative) differences in the maximum acceptable excess energies due to the particular value of the effective mass of electrons in each semiconductor. However, this influence is not expected to be pronounced since the electron effective masses (which determine the densities of states close to the bottom of the conduction band) do not differ much among common binary direct-gap semiconductors like GaAs (0.067 [14]), InP (0.08 [14]), CdTe (0.096 [25], 0.09 [26]), CdS (0.166 [26], 0.21 [27]), CdSe (0.114 [26], 0.13 [27]), ZnTe (0.11 [25], 0.12 [26]), etc. In the parentheses we included the relevant effective electron mass at the bottom of the conduction band expressed in free electron mass units. In this context, the observed virtually vanishing EHP expansion in InP at room temperature (Fig. 2) leads us to the conclusion that the related excess energy of $\sim$0.2 eV is a rough maximum for the observation of the effect in the above cited semiconductors at the pump level of the order of 10 mJ/cm$^2$.

For 800-nm pulses of Ti:sapphire lasers and among common semiconductors, the excess energy conditions is satisfied in GaAs, InP and CdTe only.

It is interesting to notice that apart from the above conditions, the developed interpretation does not involve any details about the band structure of the particular material. One crucial parameter in the incoherent regime [Eq. (2)] is the carrier effective cooling time $\tau$ which is related to the optical phonon emission rate. This parameter is not expected to have a huge spread among common semiconductors. The other fundamental parameter is the product $l_0 N_1$ which, in fact, expresses the areal density of the degenerate electron-hole gas. This product is thus given by the density of photons in the excitation pulse (provided that there is no significant loss channel like, e.g., picosecond carrier lifetime).

All these arguments suggest that the EHP expansion is controlled by the laser pulse fluence and its wavelength (through the excess energy of photoexcited charge carriers), with little differences among individual semiconducting materials.

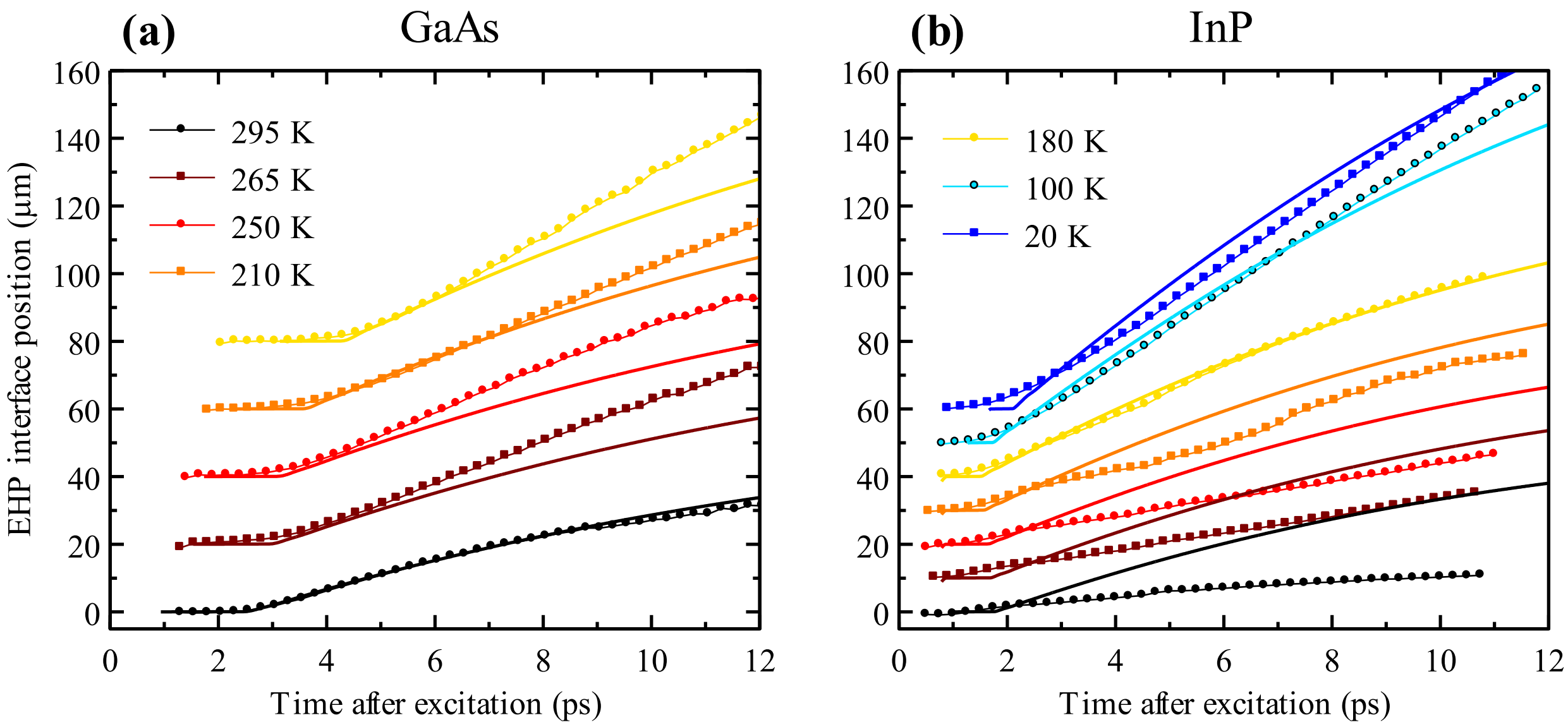

**Figure 6**: Comparison of experimental data and results of the model for incoherent regime at several temperatures for GaAs (a) and InP (b). The color code is the same for both plots; the pump fluence was $1.7 \times 10^{16}$ cm$^{-2}$ in all the cases. Vertical offset of the data was applied for clarity. Since the theory predicts shorter plateaus at the beginning of the dynamics than experimentally observed, we shifted the theoretical curves in time to match the onset of expansion with the experiment.

## 5. Conclusion

We studied the dynamics of electron-hole plasma (EHP) after strong photoexcitation in two common semiconductors, InP and GaAs, by means of ultrafast THz spectroscopy. In both materials we observed ultrafast propagation of EHP front towards the bulk of the material over distances exceeding 100 $\mu$m at low temperatures. Two mechanisms of the propagation, incoherent and coherent, were observed depending on the temperature and photocarrier excess energy, and the crossover between the two propagation regimes was experimentally identified by changing the sample temperature and the strength of optical excitation. The theoretical model for the incoherent regime introduces the notion of a carrier

“reservoir” (hot carriers in Γ-valley and two-photon absorption carriers in the *L*-valley) and an “active level” exhibiting inversion of population close to the band edge. The discussion how to correctly cast the continuous semiconductor band structure into an essentially two-level system (involving the frequently neglected fine effects like bandgap renormalization due to free carriers) is the key for a quantitative understanding of the observed dynamics as a function of temperature in the two materials. The results clearly suggest that the effect should be present in most direct bandgap semiconductors when strongly excited with low excess energy.

We also note that recently the reflection of broadband THz pulses on a propagating plasma front due to strong photoexcitation in indirect bandgap silicon was proposed as a tool for Doppler-effect-based up-conversion of THz bandwidth to higher frequencies [22]. The blue shift of THz power spectrum was then demonstrated up to a factor of 2 [11]. However, in this scheme the time window for the THz reflection is very narrow and the phase shift is not uniform within the THz pulse. The present work re-opens this opportunity for direct bandgap semiconductors where the available time window is very wide allowing reflection of picosecond THz pulses under strictly constant conditions. Similar up-conversion factors were experimentally observed in our measurements (1.46 in amplitude spectrum [9], i.e., 2.1 in power spectrum).

### Research funding

This work was supported by the Czech Science Foundation (Project No. 23-05640S) and Charles University grant SVV-2023- 260720. This work was also co-financed by European Union and the Czech Ministry of Education, Youth and Sports (Project TERAFIT - CZ.02.01.01/00/22_008/0004594).

### Author contributions

All authors have accepted responsibility for the entire content of this manuscript and approved its submission.

### Conflict of interest

Authors state no conflicts of interest.

### Data availability

The datasets generated and/or analyzed during the current study are available from the corresponding author upon reasonable request.

## Appendix: Kinetic equations for the incoherent regime

In this section, we extend and more precisely specify details of the theoretical model of the incoherent regime in order to understand the temperature dependent experimental data in both GaAs and InP.

Two photon fields are considered in the model, the first one with frequency $\omega_1$ describing the absorption at the pump pulse wavelength, and the second one ($\omega_2$) describing the stimulated emission close to the bandgap edge. These fields are coupled to two model levels in the conduction band: $N_1$ describing the number of available states per unit volume ($n_1$ being the related population of carriers) connected to the pump absorption and $N_2$ describing the number of states per unit volume ($n_2$ being the carrier population) below the quasi-Fermi level shifted inside the conduction band. The EHP expansion then corresponds to a spontaneous/stimulated emission of photons due to the recombination of carriers between the level $N_2$ and the ground state, and their reabsorption deeper in the semiconductor wafer. Two-photon absorption (TPA) was originally proposed [9] only as a mechanism leading to the saturation of the initial plasma thickness $l_0$ at high pump fluences but the electrons created by TPA were not considered to take part in any further stage of the plasma dynamics.

In the light of recent measurements [28,29,30] and theoretical calculations [31] of phonon-assisted electron relaxation, we have to admit that the *L*-valley carriers, excited by two-photon absorption, should be accounted for since their relaxation time to the $\Gamma$-valley on the picosecond time scale is fast enough to contribute to the plasma expansion dynamics. The full set of kinetic equations then reads:

$$\frac{\partial n_1}{\partial t} = \frac{\alpha}{\hbar\omega_1}\left[1 - \frac{2n_1}{N_1}\right] I_1 - \frac{1}{\tau}\left[1 - \frac{n_2}{N_2}\right] n_1 + \frac{1}{\tau_{\mathrm{L}\to\Gamma}} n_L \tag{A1}$$

$$\frac{\partial n_L}{\partial t} = \frac{\beta}{2\hbar\omega_1} I_1^2 - \frac{1}{\tau_{L\to\Gamma}} n_L \tag{A2}$$

$$\frac{\partial n_2}{\partial t} = \frac{\alpha}{\hbar\omega_2}\left[1 - \frac{2n_2}{N_2}\right] I_2 + \frac{1}{\tau}\left[1 - \frac{n_2}{N_2}\right] n_1 - A n_2 \tag{A3}$$

$$\frac{1}{v}\frac{\partial I_1}{\partial t} + \frac{\partial I_1}{\partial z} = -\alpha\left[1 - \frac{2n_1}{N_1}\right] I_1 - \beta I_1^2 \tag{A4}$$

$$\frac{1}{v}\frac{\partial I_2^{\pm}}{\partial t} \pm \frac{1}{2}\frac{\partial I_2^{\pm}}{\partial z} = -\alpha\left[1 - \frac{2n_2}{N_2}\right] I_2^{\pm} + \frac{1}{2} A\hbar\omega_2 n_2 \tag{A5}$$

Compared to [9], we added the third level denoted as "*L*" which represents the *L*-valley electrons, generated by a two-photon process (the saturation of this level is neglected since the higher excess energy and higher effective electron mass ensure that there are much more states than in the $\Gamma$-valley). The electron relaxation time from the *L*-valley to the level 1 is of the order of 0.2 ps [30]. The corresponding holes relax to the $\Gamma$-valley through a cascade of scattering events involving phonon emission and this process can last for $\sim 2$ ps [32]; consequently, we set $\tau_{L\to\Gamma} = 2$ ps. The carrier densities ($n_1$, $n_2$, $n_L$) and light intensities ($I_1$, $I_2$) are time and *z*-dependent. The sample can be divided to a part which is initially excited (by the field $\omega_1$), region A, and to an initially unexcited part, into which the plasma expands (due to the absorption of the field $\omega_2$), region B, as schematically shown in Fig. 5.

The values of the parameters of the model are as follows: $v = c/n_g$ where $c$ is the vacuum speed of light and $n_g \approx 4.2$ (GaAs) and $n_g \approx 3.6$ (InP) are the group refractive indices near the band edge, values of $N_1$ were taken from Table 1. We set $\beta = 220\ \mathrm{cm/GW}$ for both GaAs [9,10] and InP. The linear absorption coefficient $\alpha = 1.3\ \mu\mathrm{m}^{-1}$ for GaAs and $\alpha = 3.3\ \mu\mathrm{m}^{-1}$ for InP [13,33] and the spontaneous emission rate $A = 1\ \mathrm{ns}^{-1}$ [34] are known from the literature. The effective electron cooling time $\tau$ is of the order of one picosecond. The bottom level degeneracy $N_2$ appears as a crucial parameter for the dynamics and its meaning and the proper setting of its value are discussed below.

As argued in the main text, the lower limit $N_{2,c}$ of the value $N_2$ is given by the critical electron density required for the reflection of the THz radiation on the plasma edge and by the minimum density of thermalized electrons which leads to the saturation of the optical absorption in the crystal at frequency $\omega_2$. Concerning the first limitation, the THz pulse is efficiently reflected on the EHP surface if its whole spectrum lies below the electronic plasma frequency. The critical electron density for the plasma frequency at 2.5 THz (upper limit of our THz pulse spectrum) is $n_{2,0} \sim 8 \times 10^{16}\ \mathrm{cm}^{-3}$ in both InP and GaAs. This corresponds to one half of the total available number of states at the level 2, i.e., we define the first limit: $N_{2,0} = 1.6 \times 10^{17}\ \mathrm{cm}^{-3}$. The second limitation of $N_2$ comes from the requirement that the thermalized plasma does not absorb light at $\omega_2$, which enables the spatial shift of the EHP surface to the unexcited part of the crystal. To estimate this limit, we consider that the electron and hole densities are the same in the region B of the sample, that they are thermalized to the lattice temperature and that $\hbar\omega_2$ is close to the band gap energy. The saturation of absorption at $\omega_2$ arises when the population inversion at the band edge vanishes; in other words, $\mu_e - \mu_h = 0$, where the electron and hole chemical potentials $\mu_e$ and $\mu_h$ are defined with their origins at the respective band edges. This condition yields the sought electron density $n_{2,T}$ and, consequently, $N_{2,T}$ which is twice that value. The critical degeneracy $N_{2,c}$ is then the larger of the values $N_{2,0}$ and $N_{2,T}$.

Furthermore, the presence of the high-density EHP induces a band gap renormalization; its magnitude is comparable both to excess energy and excitation bandwidth ($\sim 30\ \mathrm{meV}$ for the free carrier density of $10^{18}\ \mathrm{cm}^{-3}$), and therefore it should be taken into account [35]. The band gap renormalizes differently in the initially excited region (stronger renormalization) and initially unexcited part (weaker renormalization) of the sample due to the large difference in the carrier concentration. It could then happen that all (or many of) the states of the level 2 in the region A lie below the band gap of the region B. In this case, photons emitted from the level 2 would not be (or would be weakly) absorbed in the plasma expansion region B. To ensure the occupancy of $n_{2,c}$ in this region, we have to increase the degeneracy of the level 2 by increasing its upper bound (i.e., by increasing the energy which separates the states of the levels 1 and 2).

For the calculation of the bandgap renormalization effects we use the values published in [35]. We denote by $2E_B$ the bandgap energy shrinkage in region B due to the occupancy $n_{2,c}$. The initial band gap renormalization in the region A due to an electron and hole populations averaged over this region is then denoted $2E_A$. These renormalizations evolve in time; therefore, we consider only one half of these renormalizations as effective values, i.e., $E_B$ and $E_A$, respectively (with $E_A > E_B$). It is then necessary to ensure that the photons emitted from the region A are absorbed in the region B. In order to generate the carrier density $n_{2,c}$ in region B, photons with energy $\hbar\omega_2 = \mu_e - \mu_h + E_B$ should be absorbed (here $\mu_e$ and $\mu_h$ are the chemical potentials for which the electron and the hole densities at the lattice temperature are both $n_{2,c}$). Such energy must be available within the level 2 in the region A. In other words, level 2 should involve all the states up to the electron and hole energies $\nu_e$ and $\nu_h$, respectively, that fulfill the relation $\nu_e - \nu_h + E_A = \mu_e - \mu_h + E_B$ under the assumption that the electron and hole densities are equal. This equation yields the separation energy between levels 1 and 2 in region A and thus the level degeneracy $N_2$ to be injected into the kinetic equations (A1)–(A5).